\documentclass{PoS}

\usepackage{amssymb}
\usepackage{amsmath}

\usepackage{latexsym}
\usepackage{graphicx}

\usepackage[labelformat=empty]{caption}

\newcommand{\be}{\begin{equation}}
\newcommand{\ee}{\end{equation}}

\newcommand{\dlt}{\delta}
\newcommand{\prt}{\partial}
\newcommand{\br}{{\bf r}}

\newcommand{\bt}{\beta}
\newcommand{\vp}{\varphi}
\newcommand{\ep}{\varepsilon}
\newcommand{\al}{\alpha}
\newcommand{\ra}{\rightarrow}

\newcommand{\gm}{\gamma}
\newcommand{\om}{\omega}

\newcommand{\Gm}{\Gamma}

\newcommand{\lbd}{\lambda}

\newcommand{\rgl}{\rangle}
\newcommand{\lgl}{\langle}

\title{Phase transition in multicomponent field theory at finite temperature}

\ShortTitle{Phase transition in multicomponent field theory at finite temperature}

\author{\speaker{Vyacheslav I. Yukalov}  \\
        JINR \\
        E-mail \email{yukalov@theor.jinr.ru}}

\author{Elizaveta P. Yukalova \\
        JINR \\
        E-mail \email{yukalova@theor.jinr.ru}}

\abstract{
Nuclear matter at finite temperature and barion density exhibits several
phase transitions that could happen at the early stages of the Universe
evolution and could be realized in heavy-ion or hadron-hadron collisions.
Microscopic description of phase transitions is notoriously difficult
because of the absence of small parameters. Here we present a general
approach allowing to treat situations, when there are no small parameters.
The approach is based on optimized perturbation theory and self-similar
approximation theory. It allows, starting with divergent perturbation
series in powers of an asymptotically small parameter, to construct
expressions extrapolating asymptotic series to arbitrary values of the
parameter, including its infinite limit. Examples of such approximants are:
right root approximants, left root approximants, continued root approximants,
exponential approximants, and factor approximants. The approach is
illustrated by the phase transition of gauge symmetry breaking in a
multicomponent field theory. The found critical indices are in very good
agreement with Monte Carlo simulations as well as with complicated methods
of Pad\'{e}-Borel summation, while our approach is much simpler. The nice
feature of the approach is that it gives exact values for the cases where
exact solutions are known.  }

\FullConference{XXII International Baldin Seminar on High Energy Physics
Problems,\\
                15-20 September 2014\\
                JINR, Dubna, Russia}

\begin{document}

\section{Introduction}

Varying temperature and baryon density, it is possible to realize several
different phases of nuclear matter. A qualitative phase portrait of admissible
phases, on temperature-baryon density plane, is shown in Fig. 1, where
$\rho_0 = 0.167$ fm$^{-3}$ is the normal baryon density (e.g., [1]).
The variation of temperature and baryon density can be achieved in hadron-hadron
and heavy-ion collisions. The corresponding values of these thermodynamic
variables could also exist at the early stages of the Universe evolution or in
the cores of neutron stars. There can occur phase transitions of first order,
second order, as well as crossovers [2-4].

\vskip 3mm
\begin{figure}[ht]
\centerline{\includegraphics[width=9cm]{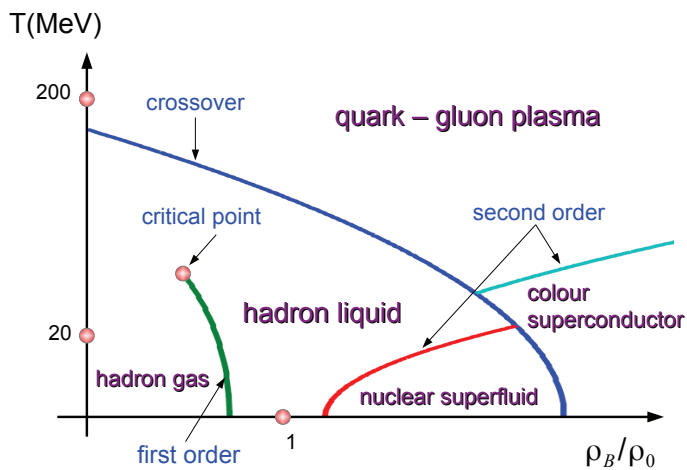}}
\caption{{\bf Figure 1}: A qualitative phase portrait of admissible
phases, on temperature-baryon density plane.
}
\label{fig:Fig.1}
\end{figure}

Phase transitions are known to be difficult for description because of the
absence of small parameters in the transition region. This especially concerns
the description of phase transitions in microscopic theory, where calculations
are possible only by resorting to a kind of perturbation theory. However,
perturbation theory usually results in divergent series that, in the best case,
have meaning in the limit of asymptotically small parameters, while the physical
parameters could be rather large, or even infinite. The notorious question is:
How it would be possible to extract information from perturbative series,
derived for asymptotically small parameters, for the values of finite and large
parameters?

There exist methods for effective summation of divergent series, such as Pad\'e
summation [5] and Borel summation [6]. However, the former exhibits a number of
deficiencies, including the appearance of spurious poles, while the second is quite
complicated and requiring the knowledge of large-order terms in power law
expansions. Both these methods not always are applicable, as is discussed in [7].

In the present report, we describe an original approach to treating divergent
perturbative series, extracting from them meaningful answers, providing good
accuracy, with being much simpler and more general than Pad\'e-Borel summation.
We illustrate the approach by the example of a phase transition in
multicomponent field theory at finite temperature.

\section{Optimized perturbation theory}

The first step of the approach is optimized perturbation theory, based on the
definition of {\it control functions} reorganizing divergent series to
convergent ones. The optimized perturbation theory was advanced in 1973, in
Thesis [8], submitted for publication in 1974, and published [9,10] in 1976.
The power of this method was illustrated by treating some anharmonic models
and strongly anharmonic quantum crystals [8-18]. Later this theory has been
applied to numerous models, under different guises and under different names,
such as modified perturbation theory, variational perturbation theory,
renormalized perturbation theory, oscillator representation, delta expansion,
optimized expansion, nonperturbative expansion, and so on (e.g., [19-24]).
All these works have used the variants of the same idea [8-18] of introducing
control functions renormalizing divergent perturbative series into convergent
series. In this section, we briefly delineate the idea of optimized perturbation
theory, as advanced in [8-18] and reviewed in [25,26].

Suppose we are interested in finding a function $f(x)$ satisfying a complicated
equation that cannot be solved exactly, but can be treated only by a kind of
perturbation theory. To explain the main idea, we consider here, for simplicity,
a real function of a real variable. A generalization to complex functions and
variables is straightforward.

Let perturbation theory give a divergent perturbative sequence $\{f_k(x)\}$,
with $k = 0,1,2,\ldots$ being an approximation-order index. The basic idea of
optimized perturbation theory is to introduce a set $\{u_k(x)\}$ of control
functions that would allow us to reorganize the divergent perturbative sequence
$\{f_k(x)\}$ into a convergent sequence $\{F_k(x,u_k)\}$, where $u_k = u_k(x)$.
A sequence is convergent if and only if it satisfies the Cauchy criterion:
For each positive $\epsilon$, there exists an order $k_\epsilon$, such that
\be
\label{1}
 | F_{k+p}(x,u_{k+p}) - F_k(x,u_k) | < \ep \;  ,
\ee
for $k > k_\epsilon$ and any positive $p = 0,1,2,\ldots$.

Control functions can be introduced in several ways, three of which are the
most common: (i) through initial conditions, (ii) through variable changes,
and (iii) through functional transformations.

\vskip 2mm

{\it Introduction of control functions through initial conditions}

\vskip 2mm

The simplest illustration of this way is when the sought function $f(x)$ is
a solution of a functional equation
\be
\label{2}
 {\cal E}[f(x) ] = 0 \;  .
\ee
Then, starting with an initial approximation $F_0(x,u)$, including a control
function $u$, it is admissible to represent the functional equation (\ref{2}) as an
iterative procedure
\be
\label{3}
   F_k(x,u_k) =  F_{k-1}(x,u_{k-1}) + {\cal E}[F_{k-1}(x,u_{k-1}) ] \; .
\ee

As a rule, physical systems are characterized by their Hamiltonians, or
Lagrangians. Say, $H$ is a Hamiltonian of a complicated system that can be
treated only by means of perturbation theory. Taking for the zero approximation
a simple Hamiltonian $H_0(u)$, containing control functions, one can rewrite
the system Hamiltonian as
\be
\label{4}
 H = H_0(u) + [ H - H_0(u) ] \;  .
\ee
Then, by perturbation theory with respect to the difference $(H - H_0)$, one
obtains higher approximations, depending on the considered problem, either for
wave functions or for Green functions. Knowing the latter, one can calculate
the corresponding approximations for observables $\hat{A}(x)$ as the averages
\be
\label{5}
 F_k(x,u_k) = \lgl \hat A(x) \rgl_k \;  ,
\ee
defined for the related $k$-order approximate wave, or Green, functions.

\vskip 2mm

{\it Introduction of control functions through variable changes}

\vskip 2mm

It is possible to change the variable $x$ through the relations
\be
\label{6}
 x = x_k(z,u_k) \; , \qquad z = z_k(x,u_k) \;  ,
\ee
involving control functions, thus, getting
\be
\label{7}
 f_k(x) = f_k(x_k(z,u_k)) \;  .
\ee
The latter expression can be expanded in powers of the new variable $z$, so
that
\be
\label{8}
 f_k(x_k(z,u_k)) \simeq \overline f_k(z,u_k) \qquad (z \ra 0 ) \;.
\ee
With the inverse variable change (\ref{6}), one has
\be
\label{9}
  \overline f_k(z,u_k)  = \overline f_k(z_k(x,u_k),u_k) \;  .
\ee
This allows us to define
\be
\label{10}
 F_k(x,u_k) = \overline f_k(z_k(x,u_k),u_k) \;  .
\ee

\vskip 2mm

{\it Introduction of control functions through functional transformations}

\vskip 2mm

The sought function can be subject to a transformation containing control
functions,
\be
\label{11}
 \hat T(u) f(x) = F(x,u) \;  ,
\ee
with the inverse transformation
\be
\label{12}
 f(x) = \hat T^{-1}(u) F(x,u) \;  .
\ee
Then we define
\be
\label{13}
  F_k(x,u_k) = \hat T(u_k) f_k(x) \; .
\ee

\vskip 2mm

{\it Formulation of equations for optimal control functions}

\vskip 2mm

After control functions are incorporated into $F_k(x,u_k)$, it is necessary
to formulate explicit equations for their calculations. By their meaning, the
control functions are to be defined in such a way that to induce convergence
for the sequence $\{F_k(x,u_k)\}$. Since convergence is characterized by the
Cauchy criterion (\ref{1}), the optimal control functions, in the spirit of optimal
control theory, can be defined as the minimizers of the Cauchy cost functional
\be
\label{14}
  {\cal C}_p[u] = \frac{1}{2} \;
\sum_k | F_{k+p}(x,u_{k+p}) - F_k(x,u_k) |^2 \; ,
\ee
whose minimization provides the fastest convergence of the sequence
$\{F_k(x,u_k)\}$. That is, we need to look for the minimal value
\be
\label{15}
\min_u | F_{k+p}(x,u_{k+p}) - F_k(x,u_k) |
\ee
of the difference $F_{k+p} - F_k$ for any given $p$.

The minimization condition (\ref{15}) involves two control functions
$u_{k+p}$ and $u_k$, which makes it impossible to define both of them
simultaneously. Hence, we cannot find the exact absolute minimum of the
Cauchy cost functional (\ref{14}), but we can try to find its approximate minimum.
Assuming that the control functions $u_{k+p}$ and $u_k$, and the related
terms $F_{k+p}$ and $F_k$ are close to each other, we can express $F_{k+p}$ as
\be
\label{16}
   F_{k+p}(x,u_{k+p}) \approx
F_{k+p}(x,u_{k}) + \frac{\prt F_k(x,u_k)}{\prt u_k} \; (u_{k+p}-u_k) \; .
\ee
Then minimization (\ref{15}) becomes
\be
\label{17}
 \min_u \left |  F_{k+p}(x,u_k) - F_k(x,u_k) +
\frac{\prt F_k(x,u_k)}{\prt u_k} \; (u_{k+p}-u_k) \right | \;  .
\ee

Depending on the relation between the difference $F_{k+p} - F_k$ and the
term containing the derivative, there can be two cases. When the derivative
term is smaller than the difference term, then minimization (\ref{17}) is
approximately satisfied under the {\it minimal difference condition}
\be
\label{18}
F_{k+p}(x,u_k) - F_k(x,u_k)  = 0 \;  .
\ee
But when the difference term is smaller than the derivative term, then
minimization (\ref{17}) is approximately valid under the condition
\be
\label{19}
 (u_{k+p}-u_k)\; \frac{\prt F_k(x,u_k)}{\prt u_k} = 0 \; .
\ee
This has to be understood as the {\it minimal derivative condition}
\be
\label{20}
\frac{\prt F_k(x,u_k)}{\prt u_k} = 0 \; ,
\ee
provided the latter possesses a solution. In case there are no solutions,
one has to set $u_{k+p} = u_k$.

As is evident, the minimal difference and minimal derivative conditions
are absolutely equivalent. Of course, in particular cases, one of them can
yield a better accuracy than the other. However, in general, it is impossible
to conclude that one is preferable to the other.

In this way, the optimized perturbation theory [8-18] consists of the
following steps. The divergent perturbative sequence $\{f_k(x)\}$ is
reorganized into the sequence $\{F_k(x,u_k)\}$ incorporating control functions
$u_k = u_k(x)$ making the latter sequence convergent. Control functions can
be introduced in three ways, through initial conditions, through variable
changes, or through functional transformations. Explicit equations for the
control functions are derived from the minimization of the Cauchy cost
functional. Approximate minimization can be done by means of either minimal
difference or minimal derivative conditions.

\section{Self-similar approximation theory}

Optimized perturbation theory has been used for various physical systems.
It is not our aim here to give a review if these numerous applications.
Just let us mention a couple of review-type articles [25,26], where further
citations can be found. Despite a variety of very successful and wide
applications of optimized perturbation theory, several questions remained
unanswered:

\vskip 2mm

(i) How it would be possible to improve accuracy within the given number of
perturbative terms?

(ii) What is a necessary condition that the Cauchy cost functional could
reach its absolute minimum, that is zero?

(iii) How to decide which of the ways of introducing control functions would
be the best one, when there are several such ways, say, by choosing different
initial approximations?

(iv) Could it be feasible to check the stability of the calculational
procedure, when no exact solutions are available, that would allow for the
explicit comparison of these exact solutions with the obtained approximations?

(v) Is it possible to define general approximants, enjoying a fixed prescribed
structure, extrapolating the series, derived for an asymptotically small
variable $x \ra 0$, to the arbitrary values of this variable from the whole
interval $[0, \infty)$?

\vskip 2mm

All these questions have been answered in self-similar approximation theory
advanced in [27-33]. The main idea of this approach is to reformulate
perturbation theory to the language of dynamical theory, considering the
approximation order $k$ as discrete time, so that the approximation sequence
$\{F_k(x,u_k)\}$ be isomorphic to the trajectory of a cascade. Then the
effective sequence limit will correspond to the cascade fixed point. And
the control of the calculational procedure stability will be equivalent
to the analysis of the dynamical system stability.

Let us define the {\it expansion function} $x = x_k(\varphi)$ by the
{\it reonomic constraint}
\be
\label{21}
 F_0(x,u_k(x) ) = \vp \; , \qquad x = x_k(\vp) \;  .
\ee
Introduce the endomorphism
\be
\label{22}
  y_k(\vp) \equiv F_k(x_k(\vp),u_k(x_k(\vp) ) )
\ee
that, owing to constraint (\ref{21}), enjoys the {\it initial condition}
\be
\label{23}
 y_0(\vp ) = \vp \;  .
\ee
The inverse to this endomorphism is
\be
\label{24}
 F_k(x,u_k(x) ) = y_k ( F_0(x,u_k(x) ) ) \;  .
\ee

In terms of this endomorphism, the Cauchy cost functional (\ref{14}) takes the form
\be
\label{25}
  {\cal C}_p[u] = \frac{1}{2} \; \sum_k | y_{k+p}(\vp) - y_k(\vp) |^2 \; .
\ee
This functional is exactly zero, provided that
\be
\label{26}
 y_{k+p}(\vp) = y_k(\vp)
\ee
for all $k \geq 0$. In particular, for $k = 0$, we have
\be
\label{27}
  y_p(\vp) = y_0(\vp) = \vp \;  .
\ee
Combining (\ref{26}) and (\ref{27}) yields the {\it functional self-similarity} relation
\be
\label{28}
y_{k+p}(\vp) = y_k(y_p(\vp) ) \;     .
\ee
Since transformation (\ref{28}) possesses the semi-group property
$y_k \circ y_p = y_{k+p}$, it can also be called {\it group self-similarity}.

Thus, the {\it self-similarity relation} (\ref{28}) is a necessary condition for
the Cauchy cost functional to be zero. Although it is not a sufficient condition.
The family of the endomorphisms $\{y_k\}$, with the group relation (\ref{28}), forms
a dynamical system in discrete time, termed {\it cascade}. By construction,
the cascade trajectory $\{y_k(\varphi): k = 0,1,2,\ldots\}$ is bijective to the
approximation sequence $\{F_k(x,u_k(x)): k = 0,1,2,\ldots\}$.

The bijectivity of the cascade trajectory and approximation sequence means the
following. If there exists the limit
\be
\label{29}
 y^*(\vp(x) ) \equiv \lim_{k\ra\infty} y_k(\vp) \;  ,
\ee
where
\be
\label{30}
 \vp(x)  \equiv \lim_{k\ra\infty} F_0(x,u_k(x) ) \;  ,
\ee
then there also exists the limit
\be
\label{31}
 F^*(x)  \equiv \lim_{k\ra\infty} F_k(x,u_k(x) ) \;    ,
\ee
such that
\be
\label{32}
 F^*(x) = y^*(\vp(x) ) \;  .
\ee

Note that the existence of a trajectory limiting point $y^*$ guarantees the
existence of a sequence limit $F^*$, however this does not guarantee that $F^*$
necessarily corresponds to the sought function $f(x)$. Such a correspondence
is an assumption typical of calculational procedures dealing with nonlinear
problems [34,35], for which the accuracy of approximations at each step cannot
be explicitly established.

For a dynamical system, the existence of a limiting trajectory point is
equivalent to the existence of a stable fixed point. Therefore, if the cascade
trajectory $\{y_k\}$, with increasing $k$, tends to a limiting point $y^*$,
the latter is a fixed point, such that
\be
\label{33}
 y_k(y^* ) = y^* \;  .
\ee

It is more convenient to deal with a dynamical system in continuous time,
instead of a system in discrete time. This can be realized by embedding the
approximation cascade into an approximation flow,
\be
\label{34}
\{ y_k(\vp) : \; k \in \mathbb{Z}_+ \} \subset
\{ y_t(\vp) : \; t \in \mathbb{R}_+ \}  ,
\ee
with the flow trajectory passing through all points of the cascade trajectory:
\be
\label{35}
 y_t(\vp) = y_k(\vp) \qquad ( t = k ) \;  .
\ee

For the dynamical system in continuous time, it is straightforward to write
down the flow evolution equation that is the Lie equation
\be
\label{36}
 \frac{\prt}{\prt t} \; y_t(\vp) = v(y_t(\vp) ) \;  ,
\ee
where the right-hand side is the flow velocity
$$
 v(\vp) \equiv \left [ \frac{\prt}{\prt t} \; y_t(\vp) \right ]_{t=0} \;  .
$$

Integrating (\ref{36}) between a given point of the cascade trajectory
$y_k = y_k(\varphi)$ and a point $y_k^* = y_k^*(\varphi)$, we get the
evolution integral
\be
\label{37}
 \int_{y_k}^{y_k^*} \; \frac{dy}{v_k(y) } = t_k \;  ,
\ee
in which $t_k$ is the time of motion from $y_k$ to $y_k^*$, while $v_k$
is the flow velocity on this time interval.

Remembering that the cascade is embedded into the flow, the flow velocity,
near the time $t = k$, employing the Euler discretization, can be expressed
through the cascade velocity
\be
\label{38}
 v_k(\vp) = F_{k+1}(x_k,u_k) - F_k(x_k,u_k) +
(u_{k+1} - u_k) \; \frac{\prt}{\prt u_k} \; F_k(x_k,u_k) \;   ,
\ee
where $x_k = x_k(\varphi)$ and $u_k = u_k(x_k)$. Invoking the bijective
relation between $y_k$ and $F_k$, the evolution integral can be represented as
\be
\label{39}
 \int_{F_k}^{F_k^*} \; \frac{d\vp}{v_k(\vp) } = t_k \;  .
\ee
The motion time $t_k$ can be treated as an additional control function.
In the simplest cases, it can be set to one or $1/k$ or defined through
additional conditions [25,26].

If $v_k$ were zero, then $y_k^*$ would be an exact fixed point of the cascade.
Unfortunately, it is difficult to set $v_k$ zero, since it contains two, yet
unknown, control functions. But we can require the minimal possible velocity,
in that way defining the control functions $u_k = u_k(x)$ by the condition
\be
\label{40}
 \min_{u_k} \left | F_{k+1}(x,u_k) - F_k(x,u_k) +
( u_{k+1} - u_k ) \; \frac{\prt}{\prt u_k} \; F_k(x,u_k)  \right | \; .
\ee
This minimization is equivalent to minimization (\ref{17}), with $p = 1$. Analogously
to the previous consideration, an approximate minimization can be done by one
of the conditions (\ref{18}) or (\ref{19}). Condition (\ref{19}) seems to be more convenient,
which results in the velocity
$$
v_k(\vp) = F_{k+1}(x_k,u_k) - F_k(x_k,u_k) \;   .
$$
Employing this in the evolution integral (\ref{39}) yields the renormalized
approximant $F_k^*$.

Defining the control functions from the minimization of the cascade velocity
implies that $y_k^*(\varphi)$ is an approximate fixed point, or quasi-fixed
point. Respectively, $F_k^*(x,u_k)$ corresponds to an effective limit $f_k^*(x)$
that is named the {\it self-similar approximation} of $f(x)$. In this way,
we have the correspondence
\be
\label{41}
 y_k^*(\vp(x)) = F_k^*(x,u_k(x) ) \ra f_k^*(x) \;  .
\ee

The improvement of the accuracy of the self-similar approximation $f_k^*(x)$,
as compared to the optimized approximation $F_k(x,u_k(x))$, is due to the
following reason. The minimization of the Cauchy cost functional is equivalent
to the minimization of the cascade velocity, which gives the optimized
approximant $F_k(x,u_k(x))$. However, the cascade velocity is not exactly zero.
The Lie equation (\ref{36}) describes the motion from the given optimized approximant
$F_k(x,u_k(x))$ to the quasi-fixed point $F_k^*(x,u_k)$ that improves the
accuracy of the former approximant.

As is mentioned above, to represent the effective limit of the approximation
sequence, the fixed point has to be stable. The stability of the procedure
here coincides with the stability of motion of the dynamical system, which is
characterized by the map multiplier
\be
\label{42}
 \mu_k(\vp) \equiv \frac{\prt}{\prt\vp} \; y_k(\vp) \;  .
\ee
The motion at the point $y_k(\varphi)$ is locally stable, provided that
\be
\label{43}
   | \mu_k(y_k(\vp)) | < 1  \; .
\ee
The multiplier at the quasi-fixed point is
\be
\label{44}
 \mu_k^*(\vp) \equiv \mu_k(y_k^*(\vp )) \;   .
\ee
The quasi-fixed point is stable, when
\be
\label{45}
 | \mu_k(F_k^*(x,u_k(x) ) | < 1 \;   ,
\ee
where relation (\ref{41}) is used.

Because in the treated case, the fixed point is a function of $x$, it is
possible to consider the maximal multiplier
\be
\label{46}
  \mu_k^* \equiv \sup_\vp | \mu_k^*(\vp) | = \sup_x
|\mu_k(F_k^*(x,u_k(x) ) ) | \; .
\ee
Then we say that a quasi-fixed point is uniformly stable, when
\be
\label{47}
 | \mu_k^* | < 1 \;  .
\ee

The analysis of the procedure stability makes it possible to answer the
question on which of the procedures is preferable, when there are several
admissible procedures differing by the way of introducing control functions.
For example, it is possible to introduce control functions by different
initial approximations, as has been analyzed for anharmonic models [25].
For strongly anharmonic quantum crystals, it is possible to choose different
initial approximations, say, Hartree or Hartree-Fock [17,36,37]. Or one
can introduce control functions by different changes of variables [6].

The answer is: That procedure is preferable that is more stable, since a
more stable procedure is assumed to be faster convergent [25,26].

Finally, we give the answer to the problem whether it is feasible to construct
general expressions extrapolating the series in powers of an asymptotically
small variable $x \ra 0$ to its arbitrary values in the whole range
$x \in [0, \infty)$.

Suppose, the sought function can be found only for an asymptotically small
variable,
\be
\label{48}
f(x) \simeq f_k(x) \qquad ( x\ra 0 ) \;   ,
\ee
where it is given by the asymptotic expansion
\be
\label{49}
 f_k(x) = f_0(x) \left ( 1 + \sum_{n=1}^k a_n x^n \right ) \;  .
\ee
Such series are usually divergent for any finite $x$.

It is convenient to consider the normalized function defined by the ratio
\be
\label{50}
 \frac{f_k(x)}{f_0(x) } = 1 + \sum_{n=1}^k a_n x^n
\ee
that, by construction, satisfies the limit
$$
\lim_{x\ra 0} \;  \frac{f_k(x)}{f_0(x) } = 1 \;  .
$$

Control functions can be introduced by applying to series (\ref{50}) the method
of fractal transforms [26,38-45] defined as
$$
 F_k(x,u) = \frac{f_k(x)}{f_0(x) } \; x^u \;  .
$$
Then, following the self-similar approximation theory by accomplishing
several times the renormalization procedure, we come, depending on the
available boundary conditions, to one of the following approximants.

\vskip 2mm

{\it Right root approximants}

\be
\label{51}
 \frac{f_k^*(x)}{f_0(x) }  = \left ( \left ( \ldots ( 1 + A_1 x)^{n_1} +
A_2 x^2 \right )^{n_2} + \ldots + A_k x^k \right )^{n_k}
\ee
that can be used, when a number of terms in the large-variable expansion
$x \ra \infty$ are known. All parameters $A_i$ and $n_i$ are uniquely defined
through this expansion [26,41,43,44].

\vskip 2mm

{\it Left root approximants}

\be
\label{52}
 \frac{f_k^*(x)}{f_0(x) }  =
\left ( \left (  \left ( \ldots ( 1 + A_1 x)^2 + A_2 x^2 \right )^{3/2} +
 A_3 x^3 \right )^{4/3} + \ldots + A_k x^k \right )^{n_k} \;  ,
\ee
in which the sole power $n_k$ is defined from the large-variable behavior
$x \ra \infty$, while all parameters $A_i$ are found from the
accuracy-through-order procedure after re-expanding (\ref{52}) in powers of $x \ra 0$
and comparing this with the initial expansion (\ref{49}). We may note that (\ref{52}) is
a particular case of (\ref{51}), with
$$
n_j = \frac{j+1}{j} \qquad ( j = 1,2, \ldots , k-1) \;   ,
$$
which involves all $A_i$ in the definition of the large-variable amplitude [7,46].

\vskip 2mm

{\it Continued root approximants}

\be
\label{53}
  \frac{f_k^*(x)}{f_0(x) }  = \left (
1 + A_1 x \left ( 1 + A_2 x \ldots ( 1 + A_k x)^s \right )^s \ldots
\right )^s \;  ,
\ee
in which the power $s$ is prescribed by the large-variable behavior, while
all parameters $A_i$ are given by the accuracy-through-order procedure at
$x \ra 0$. In the particular case of $s = -1$, approximants (\ref{53}) are reduced to
continued fractions and, hence, to Pad\'{e} approximants [47].

\vskip 2mm

{\it Exponential approximants}

\be
\label{54}
  \frac{f_k^*(x)}{f_0(x) }  =
\exp ( b_1 x \; \exp ( b_2 x \ldots \exp(b_k x) ) \ldots ) \;  ,
\ee
where the control functions $b_i$ are defined by additional conditions [26,40-42],
like the minimal difference condition (\ref{18}). More elaborated variants of defining
these control functions are also possible [26], e.g.,
$$
b_n = \frac{a_n(1+a_1^2)}{n a_{n-1}(1+a_n^2)} \qquad ( n = 1,2, \ldots, k ) \;   ,
$$
where $a_n$ are the parameters of expansion (\ref{49}).

\vskip 2mm

{\it Factor approximants}

\be
\label{55}
\frac{f_k^*(x)}{f_0(x) } = \prod_{i=1}^{N_k} ( 1 + A_i x)^{n_i} \;   ,
\ee
in which
\begin{eqnarray}
\nonumber
N_k = \left \{ \begin{array}{ll}
k/2 , & ~ k = 2,4, \ldots \\
(k+1)/2 , & ~ k = 3,5,\ldots
\end{array} \right.
\end{eqnarray}
and all parameters $A_i$ and $n_i$ are defined from the re-expansion procedure
at $x \ra 0$, equating the like-order terms [48-52].

\section{Multicomponent field theory}

Let us illustrate the application of self-similar approximation theory to
describing a phase transition in $N$-component $\varphi^4$ field theory in
$d$-dimensional space. The Hamiltonian of this field theory is
\be
\label{56}
 H[\vp] = \int \left \{ \frac{1}{2} \left [ \frac{\prt\vp(x)}{\prt x} \right ]^2 +
\frac{m^2}{2} \; \vp^2(x) + \frac{\lbd}{4!}\; \vp^4(x) \right \} \; dx \;  ,
\ee
where the standard notations are employed:
$$
 \vp(x) = \{ \vp_n(x) : \; n = 1,2,\ldots,N\} \;  ,
\qquad
x = \{ x_\al : \; \al = 1,2,\ldots,d\} \;  ,
$$
$$
\vp^2(x) \equiv \sum_{n=1}^N \vp_n^2(x) \; ,
\qquad
\left [  \frac{\prt\vp(x)}{\prt x} \right ]^2 \equiv
\sum_{n=1}^N \; \sum_{\al=1}^d \left [
\frac{\prt\vp_n(x)}{\prt x_\al} \right ]^2 \; .
$$

Hamiltonian (\ref{56}) is invariant under the reflection
\be
\label{57}
 \vp_n(x) \ra - \vp_n(x) \qquad ( n = 1,2, \ldots N ) \;  ,
\ee
so that $H[-\varphi] = H[\varphi]$. This means that the statistical average
of the field is zero: $\langle \varphi \rangle = 0$. The thermodynamic
potential
\be
\label{58}
 F [ \lgl \vp \rgl ] = - T \ln {\rm Tr} e^{-\bt H[\vp]} \;  ,
\ee
where $T$ is temperature and $\beta \equiv 1/T$, is also invariant under
reflection (\ref{57}).

It turns out that there exists a critical temperature $T_c$, such that above
this temperature, the order parameter $\langle \varphi \rangle$ is zero, while
below $T_c$ it is nonzero:
\begin{eqnarray}
\label{59}
\begin{array}{ll}
 \lgl \vp(x) \rgl = 0 & ~~~~~~~ ( T > T_c ) , \\
 \lgl \vp(x) \rgl \neq 0 & ~~~~~~~ ( T < T_c ) ,
\end{array}
\end{eqnarray}
which implies that
\be
\label{60}
 F [ \lgl \vp(x) \rgl \neq 0 ] <   F [ \lgl \vp(x) \rgl = 0 ] \; ,
\ee
for $T < T_c$. The zero $\langle \varphi \rangle$ means that all
$\langle \varphi_n \rangle$ are zero. And nonzero $\langle \varphi \rangle$
assumes that at least some of $\varphi_n$ are nonzero. This phase transition
is accompanied by the inversion symmetry breaking.

The properties of thermodynamic quantities in the critical region, where
the relative temperature
\be
\label{61}
 \tau \equiv \frac{|T - T_c |}{T_c} \ra 0
\ee
is small, are characterized by the critical indices describing the behavior of
the specific heat,
$$
 C_V \propto \tau^{-\al} \;   ,
$$
order parameter
$$
 \lgl \vp \rgl  \propto \tau^{\bt} \;   ,
$$
and isothermic compressibility
$$
\kappa_T \propto \tau^{-\gm} \;    .
$$
The dependence of an external field on the order parameter, at the critical
temperature, is of the type
$$
 h \propto  |\lgl \vp \rgl |^{\dlt} \qquad ( T = T_c)  \;   .
$$
The pair correlation function, at large distance $r \equiv |{\bf r}|$, behaves
as
$$
 g(\br) \propto \frac{\exp(-r/\xi)}{r^{d-2+\eta} } \qquad
(r\ra \infty) \;  ,
$$
with the correlation length
$$
\xi \propto \tau^{-\nu} \;   .
$$
And the vertex at $T_c$ exhibits the behavior
$$
\Gm(k) \propto 1 + ck^\om \qquad (T = T_c ) \;  .
$$

Not all seven critical indices,
$\alpha, \; \beta, \; \gamma, \; \delta, \; \eta, \; \nu, \; \omega$, are
independent. There are the so called scaling relations [53], due to Griffith,
\be
\label{62}
   \al + \bt (1 + \dlt ) =2
\ee
and Widom,
\be
\label{63}
  \gm + \bt (1 - \dlt ) =0 \;  ,
\ee
from which the Rushbrook relation
\be
\label{64}
   \al + 2\bt  + \gm =2
\ee
follows. Also, the hyperscaling relations are known:
$$
\al = 2 - \nu d \; , \qquad \bt = ( d - 2 + \eta)\; \frac{\nu}{2} \; ,
$$
\be
\label{65}
 \gm = (2 - \eta) \nu \; , \qquad \dlt = \frac{d+2-\eta}{d-2+\eta} \;  .
\ee
Thus, only three critical indices, say $\eta, \; \nu, \; \omega$, can be
treated as independent, and all others can be expressed through them.

The critical indices $\eta, \; \nu, \; \omega$ can be represented [53] as
expansions in powers of the variable $\varepsilon \equiv 4 - d$, formally
valid for asymptotically small $\varepsilon \ra 0$. We extrapolate such
$\varepsilon$-expansions by means of the factor approximants (\ref{55}) and set
$\varepsilon = 1$ corresponding to $d = 3$-dimensional space. The results for
all critical indices and different $N$ are presented in the Table. The found
values are in perfect agreement with experimental results, when these are
available, and with numerical Monte Carlo calculations, as well as with
complicated Pad\'{e}-Borel summations, as discussed in [54]. It is interesting
that for the limits $N = -2$ and $N \ra \infty$ our method provides exact
known results.

\vskip 7mm

\begin{center}

{\large{\bf Table}}: Critical indices for $N$-component $\vp^4$ field theory.

\vskip 3mm

\begin{tabular}{|c|c|c|c|c|c|c|c|} \hline
$N$&   $\al$  & $\bt$   &   $\gm$ & $\dlt$ & $\eta$  & $\nu$  & $\om$ \\ \hline
-2 &  0.5     &  0.25   &  1      & 5      & 0       & 0.5    & 0.80118 \\
-1 &  0.36844 &  0.27721&  1.07713& 4.88558& 0.019441& 0.54385& 0.79246\\
 0 &  0.24005 &  0.30204&  1.15587& 4.82691& 0.029706& 0.58665& 0.78832\\
 1 &  0.11465 &  0.32509&  1.23517& 4.79947& 0.034578& 0.62854& 0.78799\\
 2 & -0.00625 &  0.34653&  1.31320& 4.78962& 0.036337& 0.66875& 0.78924\\
 3 & -0.12063 &  0.36629&  1.38805& 4.78953& 0.036353& 0.70688& 0.79103\\
 4 & -0.22663 &  0.38425&  1.45813& 4.79470& 0.035430& 0.74221& 0.79296\\
 5 & -0.32290 &  0.40033&  1.52230& 4.80254& 0.034030& 0.77430& 0.79492\\
 6 & -0.40877 &  0.41448&  1.57982& 4.81160& 0.032418& 0.80292& 0.79694\\
 7 & -0.48420 &  0.42676&  1.63068& 4.82107& 0.030739& 0.82807& 0.79918\\
 8 & -0.54969 &  0.43730&  1.67508& 4.83049& 0.029074& 0.84990& 0.80184\\
 9 & -0.60606 &  0.44627&  1.71352& 4.83962& 0.027463& 0.86869& 0.80515\\
10 & -0.65432 &  0.45386&  1.74661& 4.84836& 0.025928& 0.88477& 0.80927\\
50 & -0.98766 &  0.50182&  1.98402& 4.95364& 0.007786& 0.99589& 0.93176\\
100 & -0.89650 & 0.48334&  1.92981& 4.99264& 0.001229& 0.96550& 0.97201\\
1000 & -0.99843 & 0.49933& 1.99662& 4.99859& 0.000235& 0.99843& 0.99807\\
10000 & -0.99986 & 0.49993&1.99966& 4.99986& 0.000024& 0.99984& 0.99979\\
$\infty$ &  -1   &  0.5   &  2    &  5     &  0      & 1      &  1 \\ \hline
\end{tabular}

\end{center}

\vskip 3mm

\section{Discussion}

We have considered the challenge of defining effective limits of divergent
series by means of renormalization techniques. The necessity of this
renormalization is dictated by the frequent occurrence of such divergent
series in complicated physical problems, e.g., arising when investigating
phase transitions. The main ideas of optimized perturbation theory and
self-similar approximation theory are presented.

The name {\it self-similarity} comes from the self-similar relation (\ref{28})
that is a necessary condition for the absolute minimum of the Cauchy cost
functional (\ref{25}). The property of group self-similarity is equivalent to
renormalization group in field theory [55]. To show this, let us introduce
the variable
\be
\label{66}
 \tau_k \equiv e^k \qquad ( k = 0,1,2, \ldots ) \;  .
\ee
Instead of (\ref{22}), we can define
\be
\label{67}
 y(\tau_k,\vp) \equiv F_k(x_k(\vp) , u_k(x_k(\vp) ) \;  .
\ee
Since $\tau_0 = 1$ at $k = 0$, the initial condition (\ref{23}) becomes
\be
\label{68}
y(1,\vp) = \vp \;   .
\ee
In the place of the self-similar relation (\ref{28}), we now have
\be
\label{69}
 y(\tau_k\tau_p,\vp) = y(\tau_k,y(\tau_p,\vp) ) \;  ,
\ee
which defines a cascade. Embedding the cascade into a flow, according to (\ref{34}),
with the discrete $\tau_k$ changing to the continuous $\tau \in [0, \infty)$,
we obtain the scaling property
\be
\label{70}
 y(\mu\tau,\vp) = y(\tau,y(\mu,\vp) ) \;  .
\ee
This group property is typical of the renormalization group equations in
field theory [56]. Instead of the Lie equation (\ref{36}), we get
\be
\label{71}
 \frac{\prt y(\tau,\vp)}{\prt\ln\tau} = \bt(y(\tau,\vp ) ) \;  ,
\ee
where the right-hand side
\be
\label{72}
\bt(\vp) \equiv \left [ \frac{\prt}{\prt\tau} \; y(\tau,\vp) \right ]_{\tau=1}
\ee
is analogous to the Gell-Mann-Low function [57].

As a simple example of the application of self-similar approximation theory,
let us briefly mention the calculation of the ground-state energy level for
the one-dimensional anharmonic oscillator [58,59] with the Hamiltonian
\be
\label{73}
\hat H = - \; \frac{1}{2} \; \frac{d^2}{dx^2} + \frac{1}{2} \; x^2 +
 g x^4 \;  ,
\ee
where $g \in [0, \infty)$ is the anharmonicity, or coupling parameter. The
introduction of a control function can be done through initial conditions,
by starting perturbation theory with the Hamiltonian
\be
\label{74}
  \hat H_0 = - \; \frac{1}{2} \; \frac{d^2}{dx^2} + \frac{u^2}{2} \; x^2 \; .
\ee
The control function $u_k(g)$ is defined by the quasi-fixed point condition (\ref{20}).
To first order of optimized perturbation theory, we have the energy $E_1 = E_1(g)$.
Using in the evolution integral (\ref{39}), the cascade velocity $v_1$ and $\tau_1 = 1$,
we find for the self-similar approximation of the ground-state energy $E^* = E^*(g)$
the equation
\be
\label{75}
 \frac{4(E^*)^2-1}{4E_1^2-1} = \exp \left \{
\frac{1}{4(E^*)^2-1} \; - \;
\frac{1}{4E_1^2-1} \; - \; \frac{1}{24} \right \} \;  .
\ee
The comparison of $E^*(g)$ with numerical calculations from the direct solution
of the Schr\"{o}dinger equation [60] shows that the found self-similar
approximation $E^*(g)$ is applicable for all values of $g \in [0, \infty)$,
yielding quite accurate results, whose maximal error does not exceed $0.1 \%$.

In Sec. 4, we have illustrated the approach by calculating the critical indices
for the inversion symmetry-breaking phase transition in an $N$-component
$\varphi^4$-field theory. The results are in prefect agreement with experimental
measurements as well as with complicated numerical techniques, such as Monte Carlo
simulations or Pad\'{e}-Borel summation. The advantage of our theory, as compared
with other numerical methods, is the combination of much greater simplicity and
very good accuracy.

\end{document}